%% file: LHCP_2014_-_QG.tex
\pdfoutput=1

\documentclass[10pt]{article}
\usepackage{graphicx}
\usepackage{xspace}
\usepackage{amsmath}

\newcommand{\Madgraph}{M\scalebox{0.8}{AD}G\scalebox{0.8}{RAPH}\xspace} 
\newcommand{\Pythia}{P\scalebox{0.8}{YTHIA}\xspace}

\newcommand{\Herwig}{H\scalebox{0.8}{ERWIG++}\xspace}

\newcommand{\pt}{\ensuremath{p_{\mathrm{T}}}}

\def\Title#1{\begin{center} {\Large #1 } \end{center}}
\def\Author#1{\begin{center}{ \sc #1} \end{center}}
\def\Address#1{\begin{center}{ \it #1} \end{center}}

\newcommand\pubblock{\rightline{\begin{tabular}{l} Proceedings of the Second Annual LHCP\\ \pubnumber\\
         \pubdate  \end{tabular}}}

\newenvironment{Abstract}{\begin{quotation} \begin{center} 
             \large ABSTRACT \end{center}\bigskip 
      \begin{center}\begin{large}}{\end{large}\end{center} \end{quotation}}

\newenvironment{Presented}{\begin{quotation} \begin{center} 
             PRESENTED AT\end{center}\bigskip 
      \begin{center}\begin{large}}{\end{large}\end{center} \end{quotation}}

\input econfmacros.tex

\usepackage{color}

\newcommand\pubnumber{ CMS-CR-2014/178 }

\newcommand\pubdate{\today}

\def\affiliation{
On behalf of the CMS Collaboration, \\
Department of Elementary Particle Physics \\
Universiteit Antwerpen, 2020 Antwerp, Belgium}

\begin{document}

\large
\begin{titlepage}
\pubblock

\vfill
\Title{Quark-gluon Jet Discrimination At CMS}
\vfill

\Author{ Tom Cornelis  }
\Address{\affiliation}
\vfill
\begin{Abstract}
Many physics analyses at the LHC are looking into processes where the signal jets are originating from quarks, while jets in the background are more gluon enriched.
Based on observables sensitive to fundamental differences in the fragmentation properties of gluons and quarks, 
a likelihood discriminant is constructed to distinguish between jets originating from quarks and gluons.
The performance of the tagger is evaluated using Z+jets and dijet events produced in proton-proton collisions at a centre-of-mass energy of 8 TeV, recorded by the CMS experiment at the LHC.
\end{Abstract}
\vfill

\begin{Presented}
The Second Annual Conference\\
 on Large Hadron Collider Physics \\
Columbia University, New York, U.S.A \\ 
June 2-7, 2014
\end{Presented}
\vfill
\end{titlepage}
\def\thefootnote{\fnsymbol{footnote}}
\setcounter{footnote}{0}
%

\normalsize 

\def\plotHeight{4.75cm}
\section{Construction of a quark-gluon likelihood discriminant}
\begin{figure}[htb]
    \centering
    \includegraphics[height=\plotHeight]{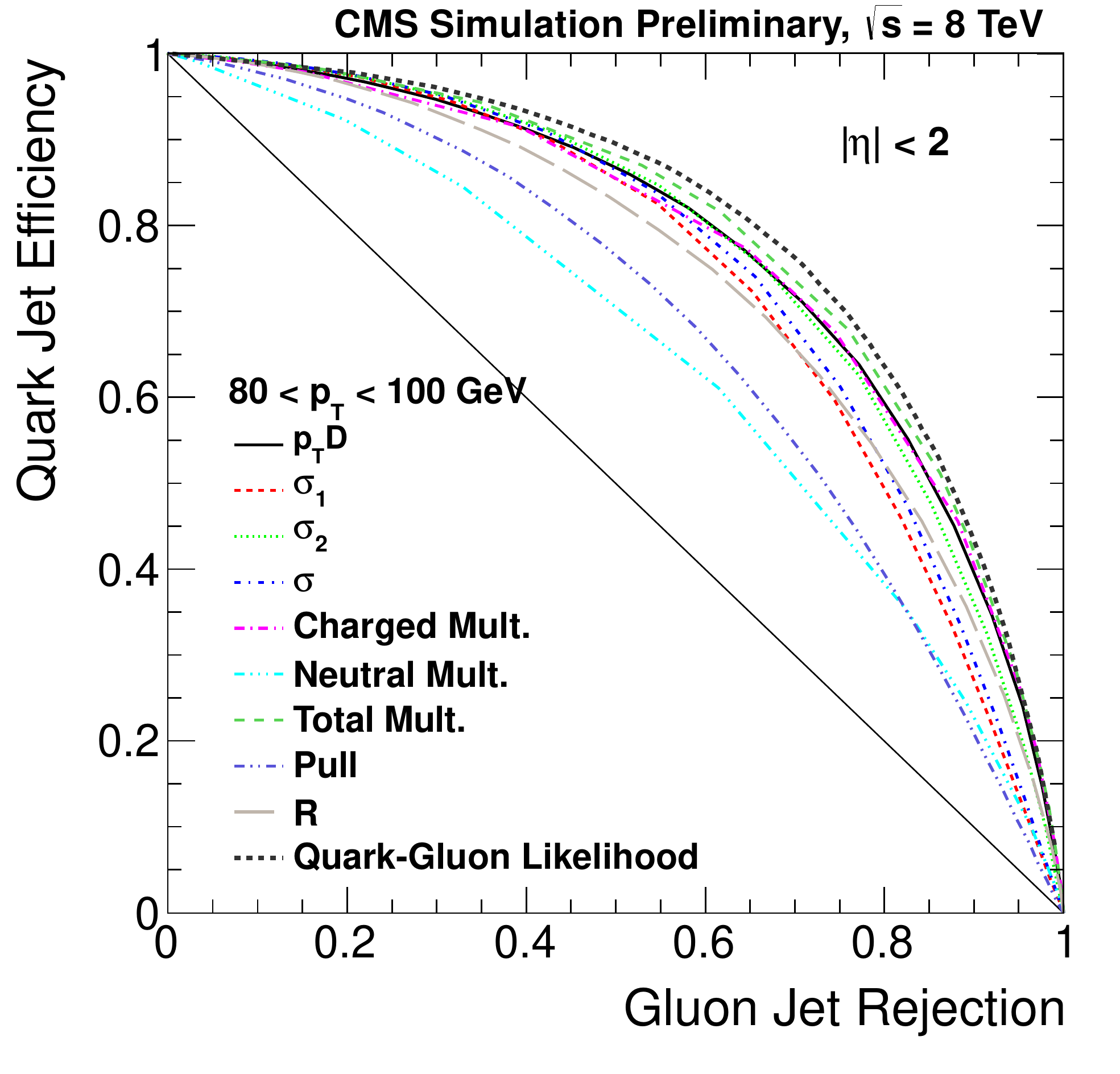}
    \includegraphics[height=\plotHeight]{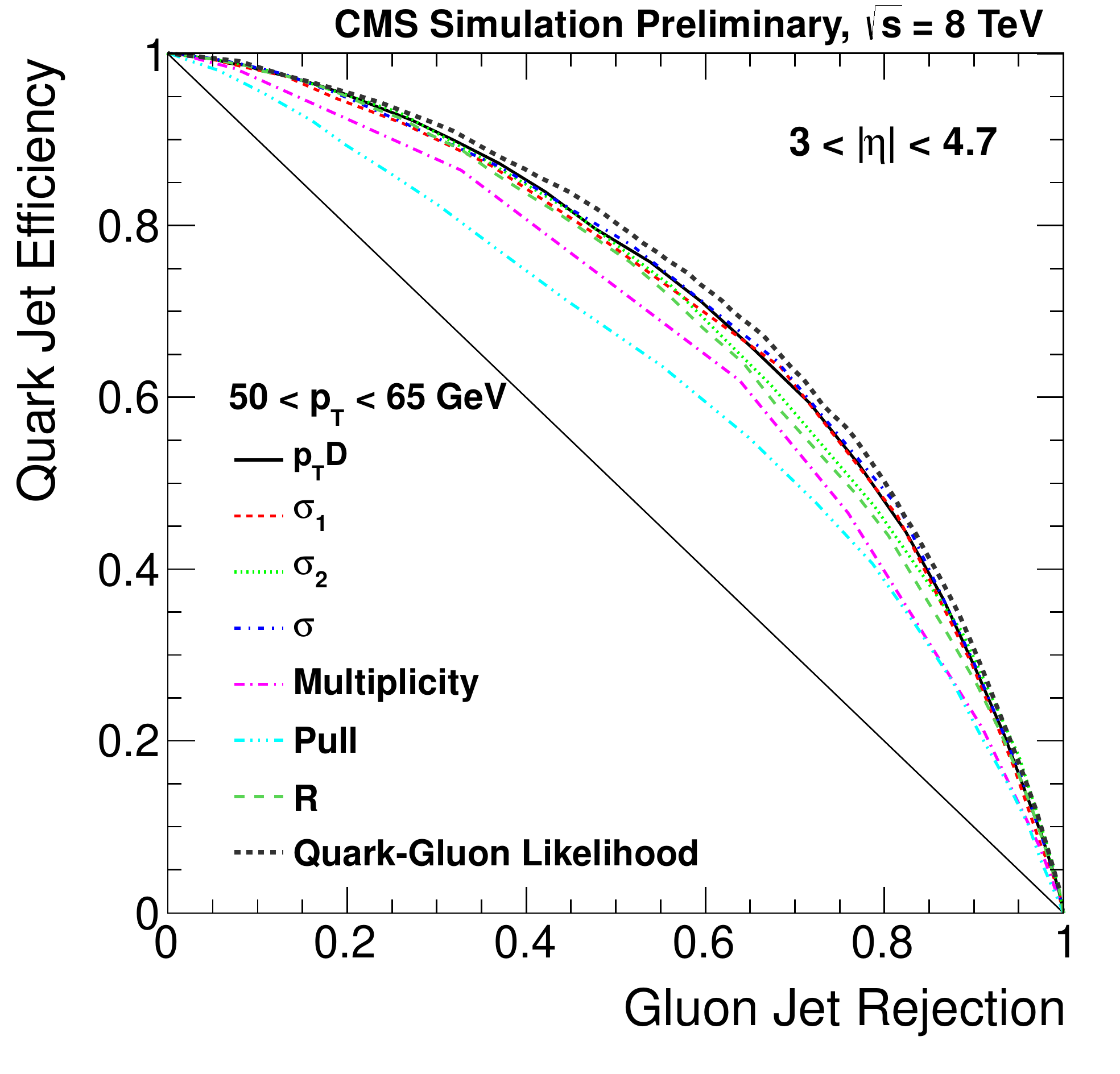}
    \caption{Single variable performance comparisons for quark-gluon jet discrimination using ROC curves for central jets with $80<\pt<100$ GeV {\it(left)} and forward jets with $50<\pt<65$ GeV {\it(right)}.
    The rationale and detailed definition of the chosen variables can be found in Ref. \cite{CMS:2013kfa}.}
    \label{fig:ROC}
  \end{figure}

  Because of different colour interaction and hadronization, gluon jets are wider, with higher multiplicities and have a more uniform energy fragmentation, while
  quark jets are more likely to produce narrow jets with hard constituents that carry a significant fraction of the energy.
  The CMS quark-gluon likelihood discriminant makes use of these jet properties through variables provided by the CMS particle-flow reconstruction. 
  The individual discrimination performances of such variables can be described using 
  receiver operating characteristic (ROC) curves (Figure \ref{fig:ROC}). These curves, derived from simulation, show the efficiency to select a quark jet 
  for possible selection cuts used to reject different fractions of gluon jets.

  Based on their performance and robustness with respect to track reconstruction, particle identification and pile-up, three variables are chosen to build a likelihood discriminator:
  \begin{itemize}
    \item the \textbf{multiplicity}, \ie\ the total number of particle flow candidates reconstructed within the jet
    \item the jet \textbf{energy sharing} variable
     \begin{equation}
      \pt D=\frac{ \sqrt{\sum_i p_{\mathrm{T},i}^2}}{\sum_i p_{\mathrm{T},i}} \nonumber
     \end{equation}
     which has $\pt D\rightarrow 1$ for jets made of only one particle that carries all of its momentum and $\pt D\rightarrow 0$ for a jet made of an infinite number of particles
    \item the \textbf{angular spread} is measured by minor axis $\sigma_2$ of the jet in the $\eta-\phi$ plane
  \end{itemize}

  \begin{figure}[htb]
    \centering
       \includegraphics[height=\plotHeight]{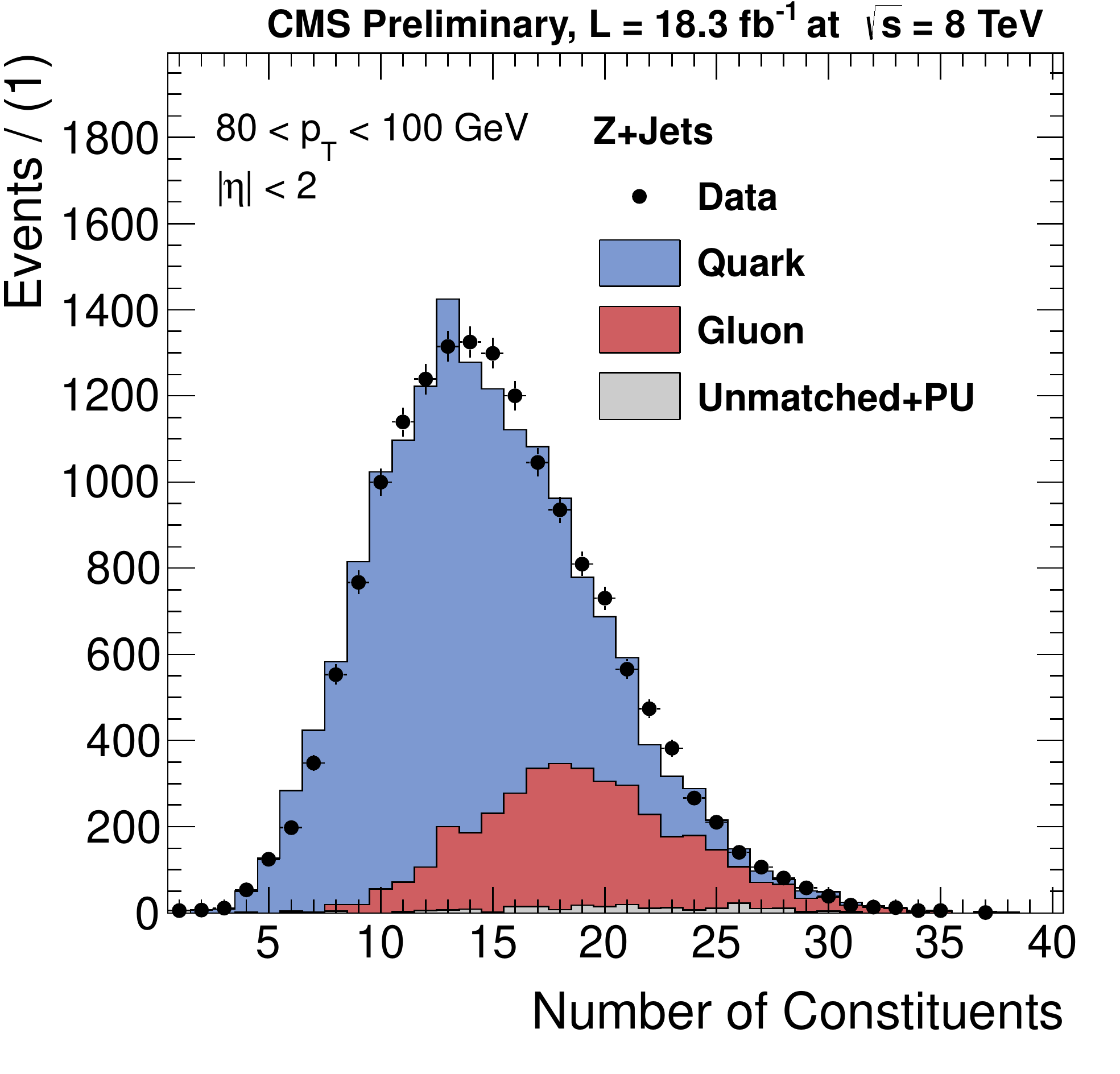}
       \includegraphics[height=\plotHeight]{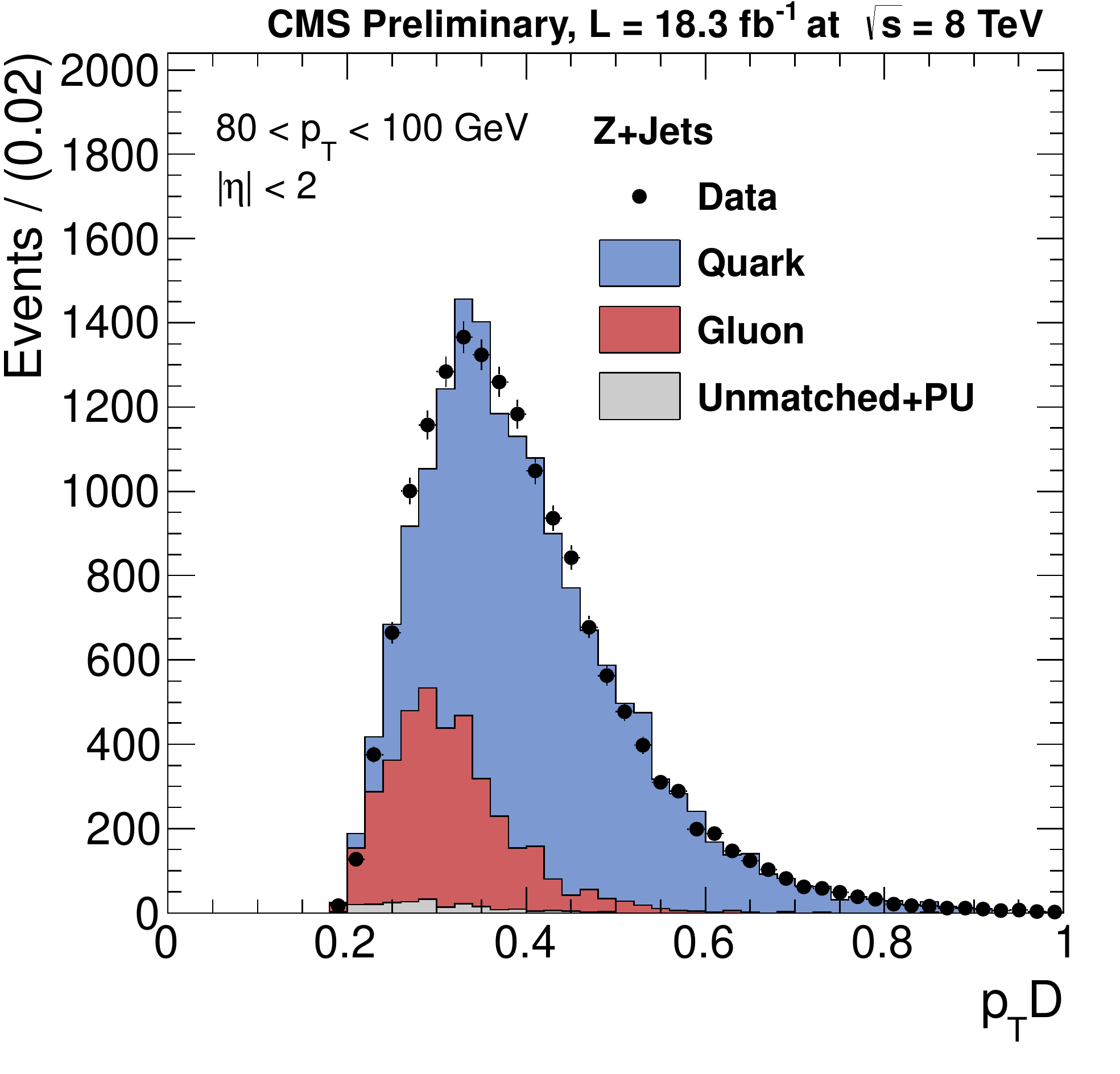}
       \includegraphics[height=\plotHeight]{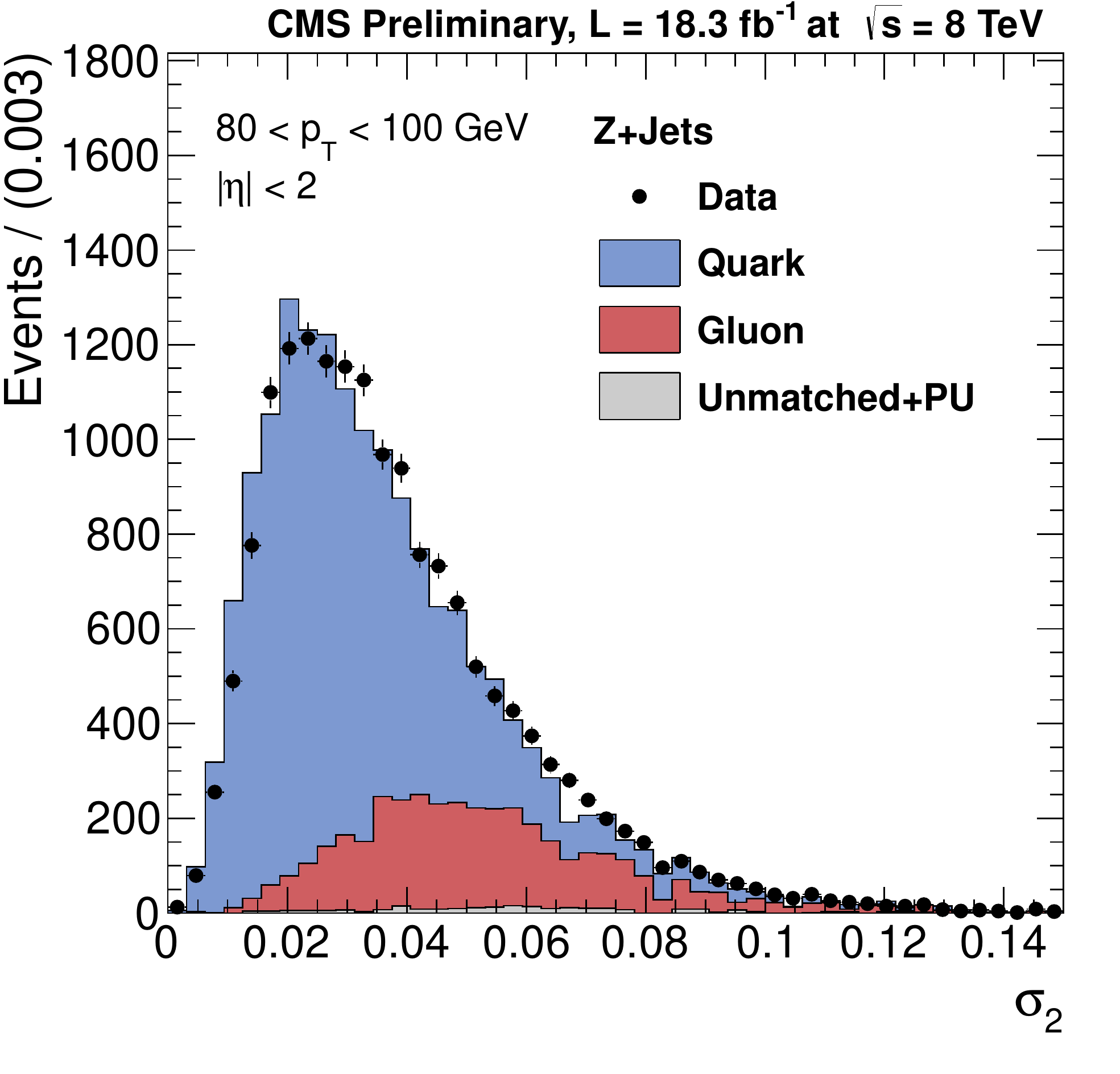}
       \caption[variables]{Comparison of data with \Madgraph + \Pythia 6 simulation, for jets with $80 < \pt < 100$ GeV and $|\eta| < 2$ in Z+jet events, for the three input variables used in the likelihood discriminator.}
     \label{fig:variables}
  \end{figure}
  
  A better discrimination power and stability to pile-up effects is found by restricting the charged particle flow candidates to those linked to tracks compatible with the 
  primary interaction vertex, and restricting the neutral particle flow candidates to those who have a transverse momentum larger than 1 GeV.
  The likelihood discriminant is binned in $\pt$ and pile-up ($\rho$) in order to account for the strong dependence of the means and shapes of the variables. The discriminant is constructed
  for jets in both the central (with pseudorapidity $|\eta| < 2.4$) and forward region (with $2.4 < |\eta| < 4.7$).
        
\section{Validation on data}
\begin{figure}[htb]
    \centering
       \includegraphics[height=\plotHeight]{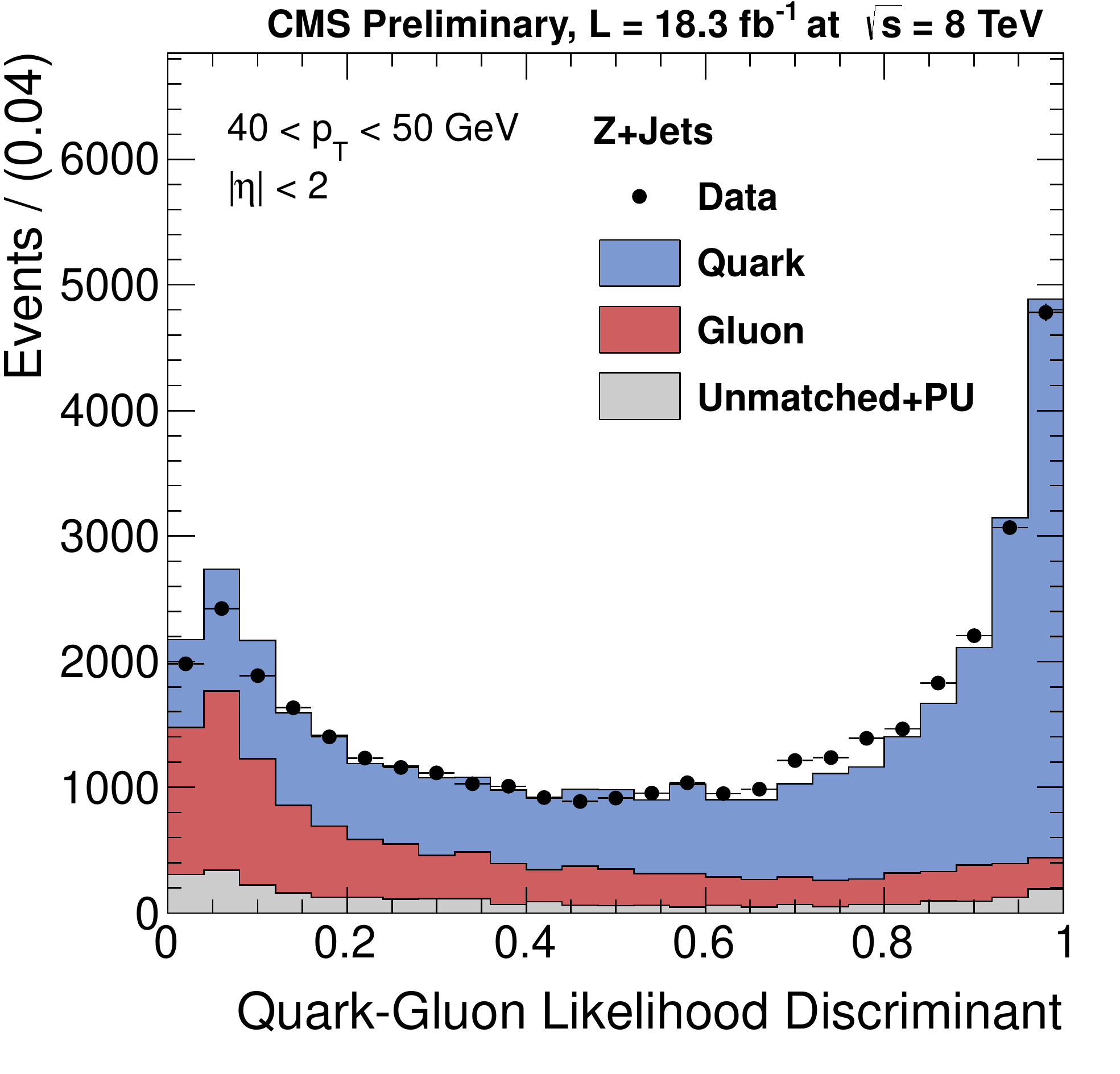}
       \includegraphics[height=\plotHeight]{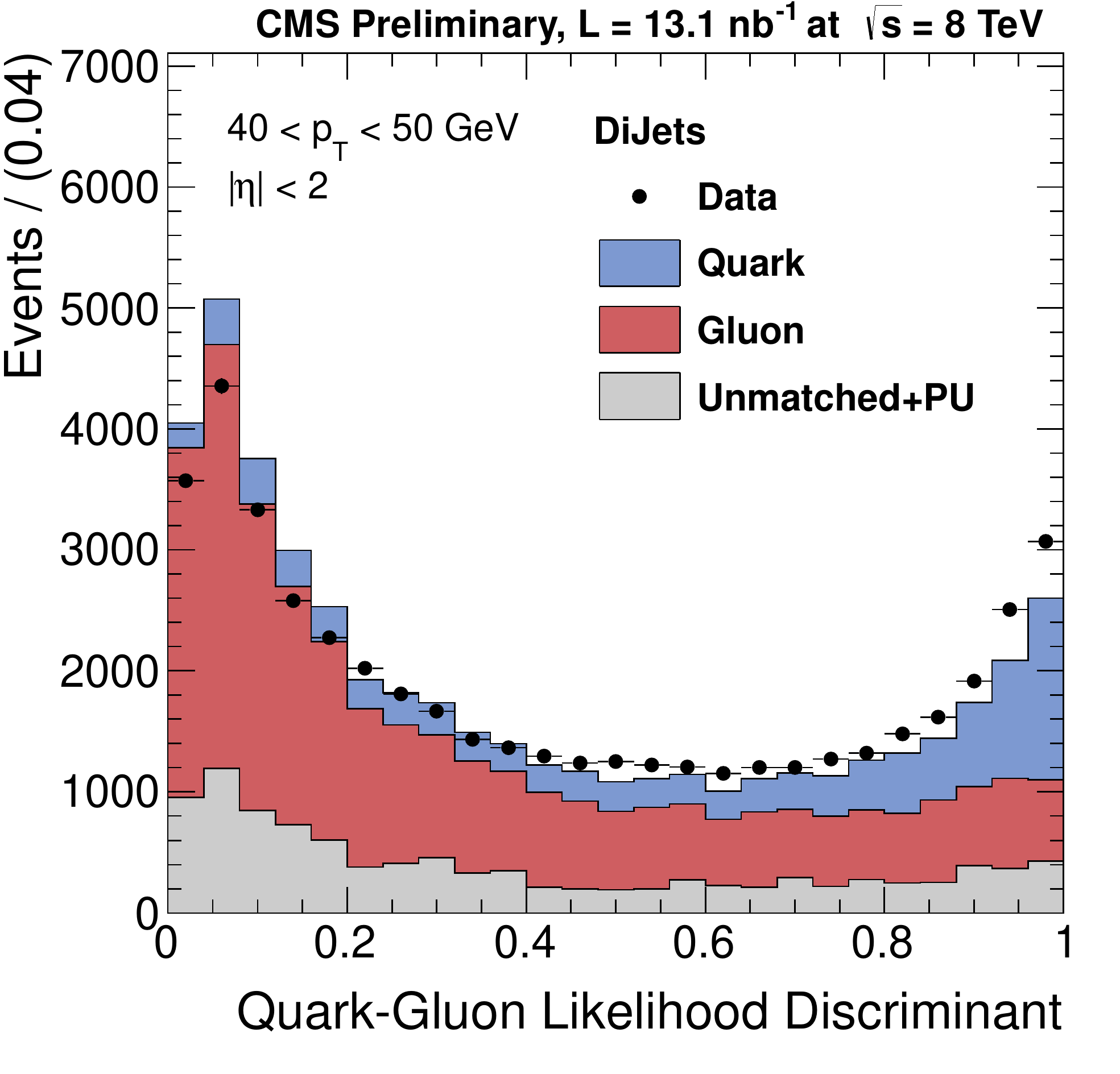}
       \includegraphics[height=\plotHeight]{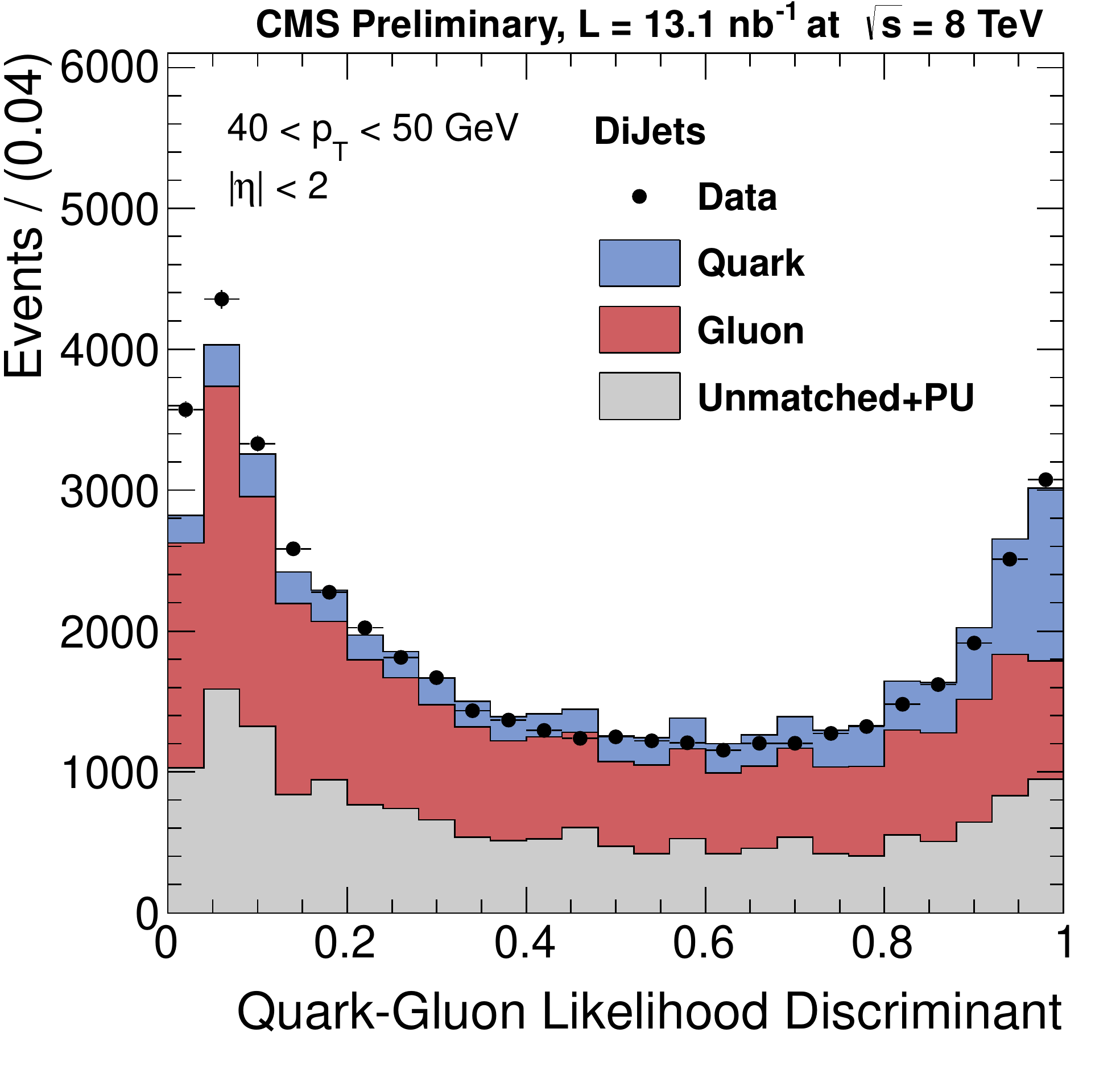}
       \caption[data validation]{Data validation of the likelihood discriminant for jets with $40 < \pt < 50$ and $|\eta| < 2$, comparing Z+jets data with \Madgraph + \Pythia 6 simulation {\it(left)}, dijet data with \Pythia 6 simulation {\it(middle)}
       and dijet data with \Herwig simulation {\it(right)}.}
     \label{fig:dataValidation}
  \end{figure}

  The performance of the discriminator has been validated on 8 TeV data by identifying two control samples, each aimed at enriching one of both parton flavours. A Z+jets control sample, with
  the leading jet being back-to-back with the Z in the transverse plane by requiring their azimutal difference to be greater than 2.5 radians, is expected to offer a relatively 
  pure sample of quark jets. A dijet sample, where the back-to-back requirement is applied on the azimutal angle between the two leading jets, provides us a gluon-enriched sample.
  Validations for both data samples, compared with simulation, are shown in Figure \ref{fig:dataValidation}.

\section{Shape uncertainty on the likelihood discriminant}  
\begin{figure}[htb]
    \centering
       \includegraphics[height=\plotHeight]{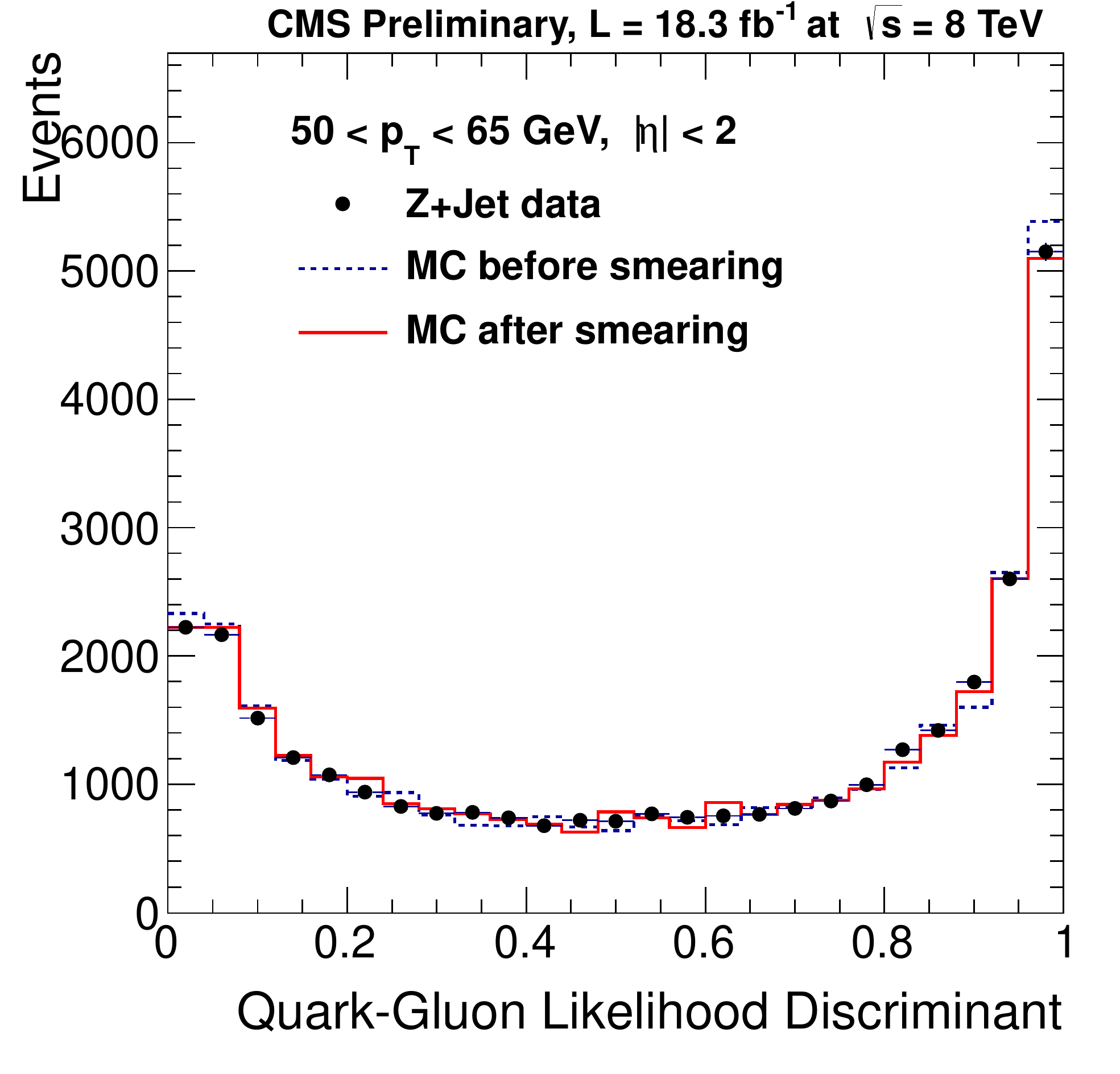}
       \includegraphics[height=\plotHeight]{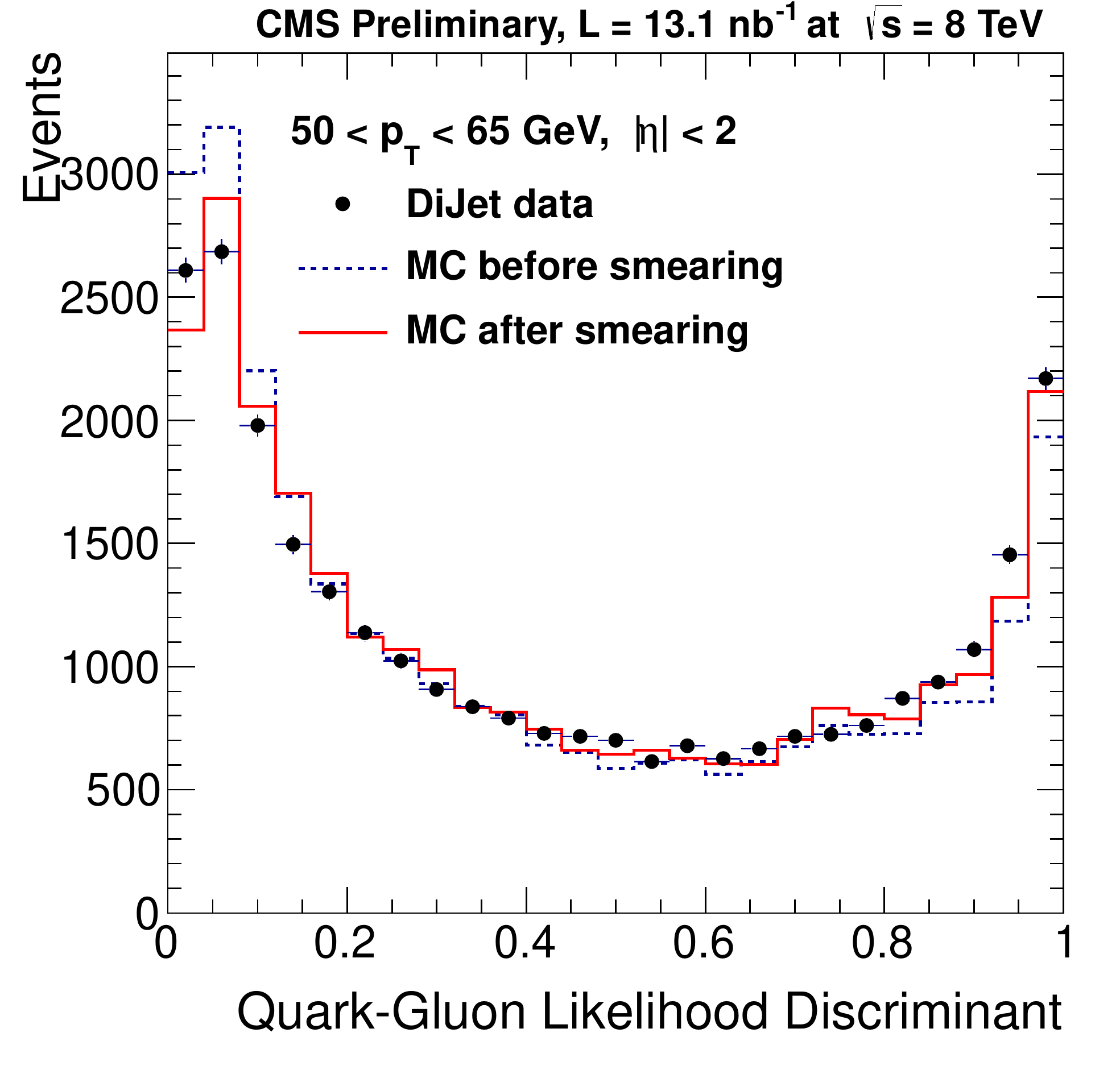}
       \caption{Validation of the smearing function method for jets with $50 < \pt < 65$ GeV and $|\eta|<2$ in dijet events. The data is compared to the simulation before and after the application of the smearing.
       The smearing function was derived in the Z+jets case {\it(left)}, good closure is observed when applied to the dijet case {\it(right)}.}
     \label{fig:smearings}
    \end{figure}

    A general applicable recipe is developed to estimate the uncertainty on the likelihood discriminant output. A smearing function is chosen to vary
    the discriminator shape in simulation to match the shape in data:
    \begin{equation}
    g\left(x,a,b\right) =  \frac{1}{2}\tanh\left(a \;\textup{arctanh}\left(2 x - 1  \right) + b\right) + \frac{1}{2} \label{eqn:syst:g} \nonumber
    \end{equation}
    The two parameters $(a,b)$ allow the population to shift through the center and towards the center or the extremes, while still keeping the distribution between 0 and 1.
    The values of these parameters are obtained by a minimisation of the $\chi^2$ obtained from a comparison between data and simulation. The same smearing functional form is
    applied independently on the quark and gluon distributions. The application of the smearing to the likelihood discriminant is shown in Figure \ref{fig:smearings}.
    
\begin{figure}[htb]
    \centering
       \includegraphics[height=\plotHeight]{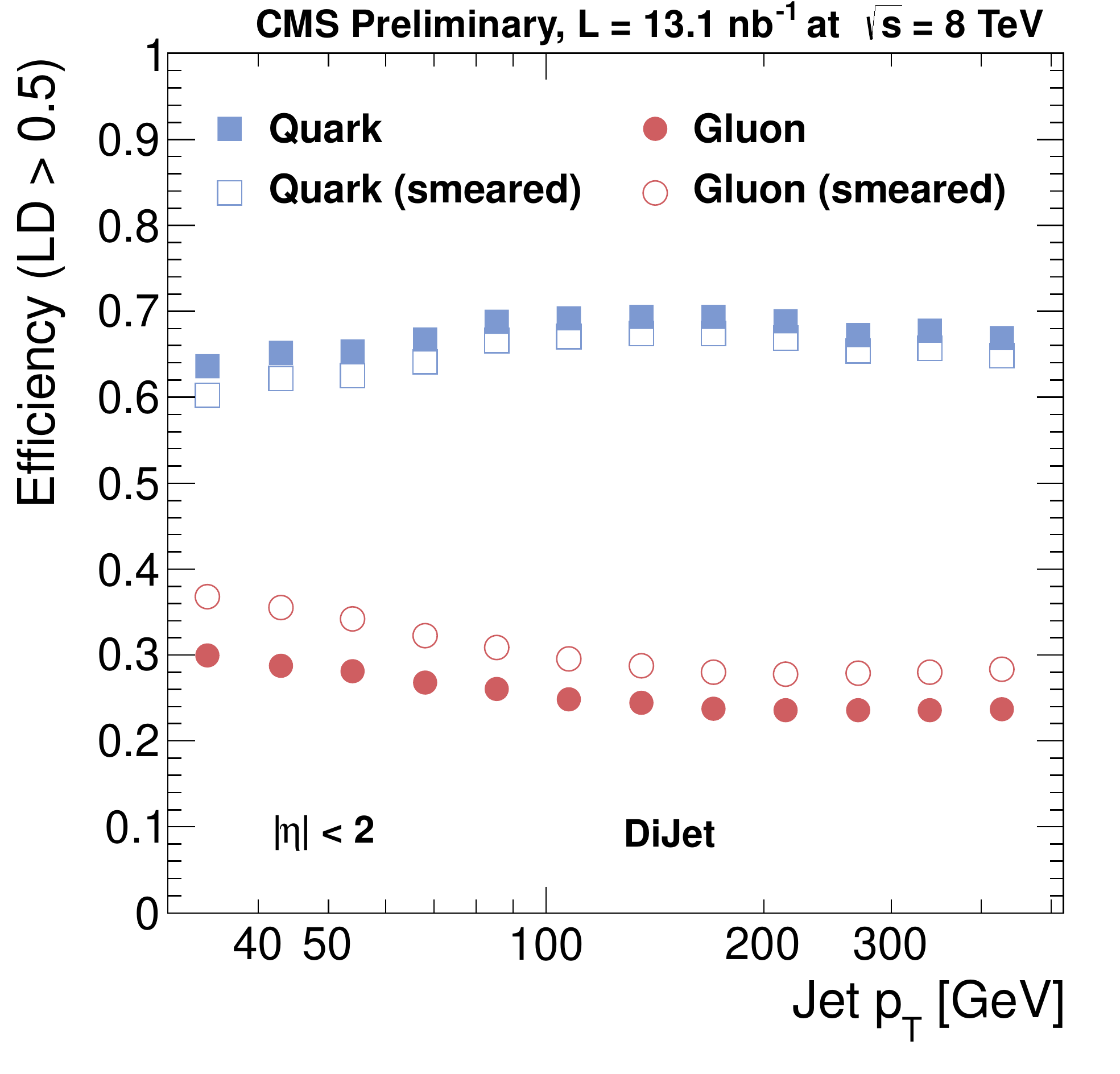}
       \includegraphics[height=\plotHeight]{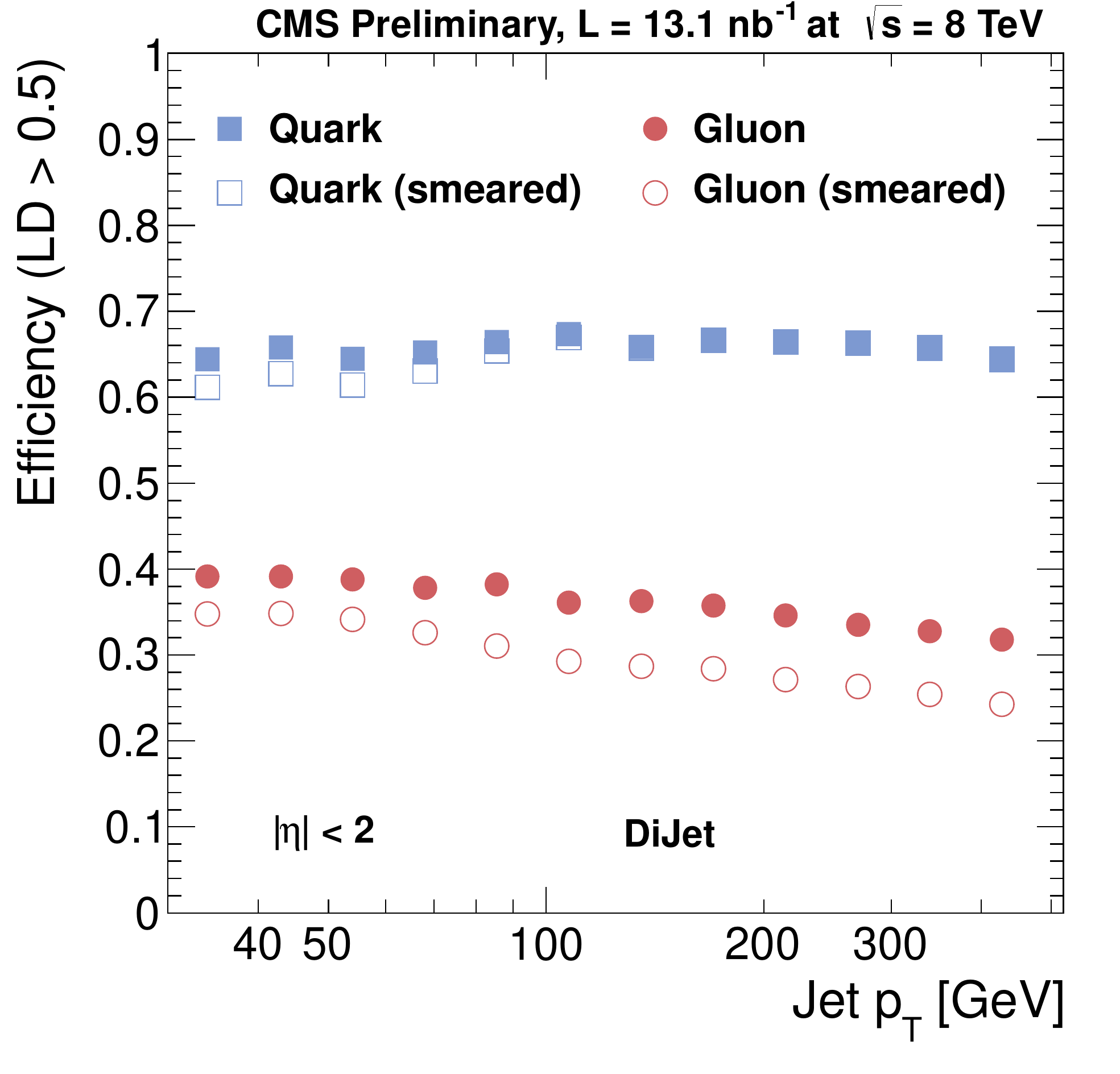}
       \includegraphics[height=\plotHeight]{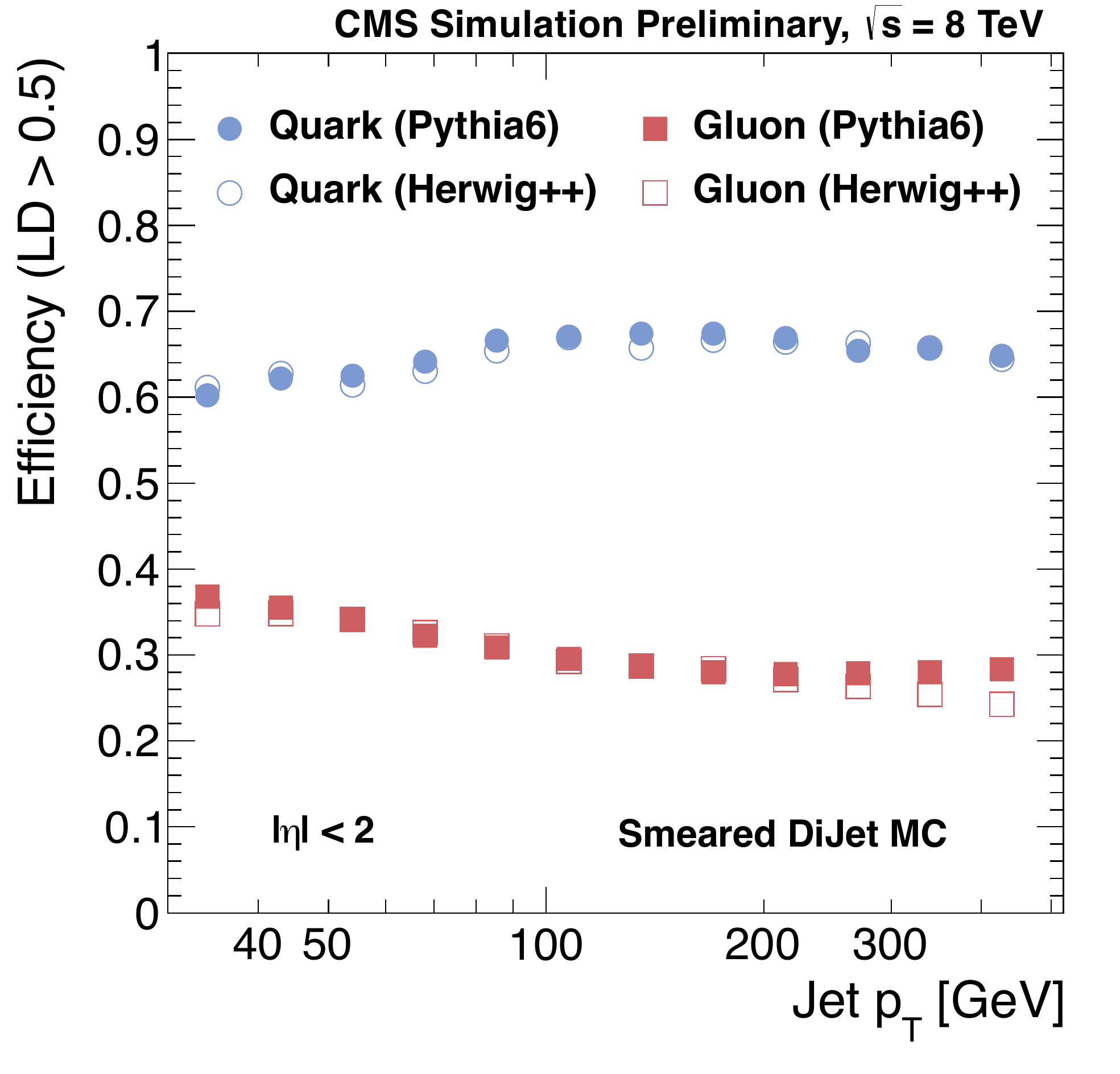}
       \caption[Change in discriminating performance by comparing the fraction of quarks and gluons with a quark-gluon likelihood discriminant 
       greater than 0.5 as a function of the jet transverse momentum, before and after smearing.]{Change in discriminating performance by comparing the fraction of quarks and gluons with a quark-gluon likelihood discriminant 
       greater than 0.5 as a function of the jet transverse momentum, before and after smearing. Different smearings were retrieved for \Pythia 6 {\it(left)} and \Herwig {\it(middle)}, 
       and are in agreement with each other {\it(right)}.}
     \label{fig:pythiaVSherwig}
    \end{figure}

    Because a different hadronization model will result in slightly different input variables, and therefore a different discriminator output, the optimisation of the smearing
    parameters is performed independently for both \Pythia 6 and \Herwig simulations. As shown in Figure \ref{fig:pythiaVSherwig}, the smearings correct for the worse discriminating performance
    in data compared to the \Pythia6 simulation, and for the better performance compared to the \Herwig simulation. After applying the smearing functions, \Pythia 6 and \Herwig
    simulations are in agreement with data, and predict the same discriminator performances.
  
\section{Conclusions}
A likelihood discriminant has been developed to separate jets originating from gluons or light-quarks. The discriminator input variables and output distributions have been validated 
using Z+jets and dijet events. A smearing function is applied to distort the shape of the output distributions in simulation, in order to reproduce better the observed data outputs.


\end{document}

%% file: econfmacros.tex
\def\beq{\begin{equation}}
\def\eeq#1{\label{#1}\end{equation}}
\def\eeqn{\end{equation}}

\def\beqa{\begin{eqnarray}}
\def\eeqa#1{\label{#1}\end{eqnarray}}
\def\eeqan{\end{eqnarray}}

\let\bar=\overbar

\def\ie{{\it i.e.}}

\def\Dslash{\not{\hbox{\kern-4pt $D$}}}
\def\dslash{\not{\hbox{\kern-2pt $\del$}}}

\def\msb{{\bar{\ssstyle M \kern -1pt S}}}

